# AI-based soundscape analysis: Jointly identifying sound sources and predicting annoyance[a)]

Yuanbo Hou,[1,b)] Qiaoqiao Ren,[2] Huizhong Zhang,[3] Andrew Mitchell,[3] Francesco Aletta,[3] Jian Kang,[3] and Dick Botteldooren[1]

[1]*Wireless, Acoustics, Environmental, and Expert Systems Research Group, Department of Information Technology, Ghent University, Gent, 9052, Belgium*

[2]*AI and Robotics, Internet Technology and Data Science Lab, Department of Electronics and Information Systems, Interuniversity Microelectronics Centre, Ghent University, Gent, 9052, Belgium*

[3]*Institute for Environmental Design and Engineering, The Bartlett, University College London, London, WC1H 0NN, United Kingdom*

**ABSTRACT:**
Soundscape studies typically attempt to capture the perception and understanding of sonic environments by surveying users. However, for long-term monitoring or assessing interventions, sound-signal-based approaches are required. To this end, most previous research focused on psycho-acoustic quantities or automatic sound recognition. Few attempts were made to include appraisal (e.g., in circumplex frameworks). This paper proposes an artificial intelligence (AI)-based dual-branch convolutional neural network with cross-attention-based fusion (DCNN-CaF) to analyze automatic soundscape characterization, including sound recognition and appraisal. Using the DeLTA dataset containing human-annotated sound source labels and perceived annoyance, the DCNN-CaF is proposed to perform sound source classification (SSC) and human-perceived annoyance rating prediction (ARP). Experimental findings indicate that (1) the proposed DCNN-CaF using loudness and Mel features outperforms the DCNN-CaF using only one of them. (2) The proposed DCNN-CaF with cross-attention fusion outperforms other typical AI-based models and soundscape-related traditional machine learning methods on the SSC and ARP tasks. (3) Correlation analysis reveals that the relationship between sound sources and annoyance is similar for humans and the proposed AI-based DCNN-CaF model. (4) Generalization tests show that the proposed model's ARP in the presence of model-unknown sound sources is consistent with expert expectations and can explain previous findings from the literature on soundscape augmentation. 

## I. INTRODUCTION

To mitigate the effect of urban sound on the health and well-being of city dwellers, previous research has classically focused on treating noise as a pollutant. For a couple of decades, researchers have gradually changed their focus to a more holistic approach to urban sound, referred to as the soundscape approach (Brambilla and Maffei, 2010; Kang *et al.*, 2016; Nilsson and Berglund, 2006; Raimbault and Dubois, 2005). Overall, the soundscape approach offers a more comprehensive understanding of urban sound and has the potential to lead to more effective interventions for improving the health and well-being of city dwellers (Abraham *et al.*, 2010; Tsaligopoulos *et al.*, 2021).

Previous studies on the categorization and quantification of soundscapes mostly rely on assessments of participant perceptions. In these studies (Acun and Yilmazer, 2018; Bruce and Davies, 2014; Mackrill *et al.*, 2013; Maristany *et al.*, 2016), participants are usually guided to participate in questionnaires about soundscapes. For example, based on investigations, Yilmazer and Acun (2018) explore the relationship among the sound factors, spatial functions, and properties of soundscapes. Using field questionnaires, Fang *et al.* (2021) explore how different participants' perceptions and preferences for soundscapes differed. Questionnaires may include a direct assessment of the soundscape quality, but the appraisal is often indicated in the two-dimensional plane spanned by pleasantness and eventfulness (Axelsson *et al.*, 2010). To benchmark soundscape emotion recognition in a valence-arousal plane, Fan *et al.* (2017) created the Emo-soundscapes dataset based on 6-s excerpts from Freesound.org and online labeling by 1182 annotators. They later used it for constructing a deep learning model for automatic classification (Fan *et al.*, 2018). To automatically recognize the eventfulness and pleasantness of the soundscape, Fan *et al.* (2015) builds a gold standard model and tests the correlation between the level of pleasure and the level of eventfulness. In an

[a)]This paper is part of a special issue on Advances in Soundscape: Emerging Trends and Challenges in Research and Practice.

[b)]Email: Yuanbo.Hou@UGent.be

everyday context, uneventful and unpleasant soundscapes are often not noticed and do not contribute to the experience of the place. Hence, Sun *et al.* (2019) propose a soundscape classification that acknowledges that sonic environments can be pushed into the background. Only foregrounded soundscapes contribute to the appraisal and are classified as disruptive and supportive, the latter being either calming or stimulating (Sun *et al.*, 2019). Based on audio recordings containing implicit information in soundscapes, Thorogood *et al.* (2016) established the background and foreground classification task within a musicological and soundscape context. For urban park soundscapes, Gong *et al.* (2022) introduce the concepts of "importance" and "performance" and position the soundscape elements in this two-dimensional plane. The importance dimension reflects to what extent a particular sound is an essential part of this soundscape. The perception study underlying this paper (Mitchell *et al.*, 2022) can be seen as a foregrounded soundscape assessment with annoyance as its primary dimension, which is a negative dimension of soundscape assessment. This type of assessment is often used to identify sources of noise, and it allows researchers to identify sources of annoyance that can cause negative health reflections.

Research on annoyance has been carried out based on non-acoustic approaches from different perspectives in the fields of psychology, sociology, medicine, human-computer interaction, and vision. In psychology, researchers primarily focus on the effects of emotion and mood on annoyance (Timmons *et al.*, 2023). The findings of the DEBATS study (Lefèvre *et al.*, 2020) also confirm that considering non-acoustic factors such as situational, personal, and attitudinal factors will improve annoyance predictions. Sociological studies tend to pay more attention to the impact of social support, social relationships, and cultural factors on annoyance (Beyer *et al.*, 2017). In medical studies, the relationship between annoyance and health is emphasised (Eek *et al.*, 2010). The study of Carlsson *et al.* (2005) indicates that the correlation between subjective health and functional ability increases with increasing annoyance levels. Human-computer interaction usually utilises user experience studies, visual eye tracking, and virtual reality techniques to analyse and predict the annoyance of users when interacting with machines (Mount *et al.*, 2012). On the other hand, acoustic-based annoyance research focuses more on the effects of sound and auditory stimulation on an individual's psychological and emotional state (Nering *et al.*, 2020). The relationship between the appraisal of the soundscape and the assessment of annoyance on the community level is still under-researched, although it was first explored in 2003 (Lercher and Schulte-Fortkamp, 2003). Several non-acoustic factors influence community noise annoyance, and some of them, such as noise sensitivity (Das *et al.*, 2021) are so strongly rooted in human auditory perception (Kliuchko *et al.*, 2016) that they probably also contribute to soundscape appraisal.

The formal definition of "soundscape" refers to an understanding of the sonic environment, hence recognizing sources. The influence of perceived sounds on the appraisal of soundscapes is found to depend on contexts (Hong and Jeon, 2015). The sounds that people hear in their environment can have a substantial impact on their overall appraisal of that environment, and it can influence people's emotional and cognitive responses to their living surroundings. Hence, any acoustic signal processing attempting to predict this, is very likely to benefit from automatic sound recognition. Automatic sound recognition in predicting people's perception of the soundscape has the potential to improve our understanding of how the acoustic environment affects our perceptions, and can inform the development of more effective interventions to promote positive outcomes. Therefore, Boes *et al.* (2018) propose to use an artificial neural network to predict both the sound source (human, natural, and mechanical) perceived by users of public parks as well as their appraisal of the soundscape quality. It was shown that sound recognition outperforms psychoacoustic indicators in predicting each of these perceptual outcomes. Identifying specific sounds in urban soundscapes is relevant for assisting drivers or self-driving cars (Marchegiani and Posner, 2017), and for urban planning and environment improvement (Ma *et al.*, 2021).

In this paper, a new artificial intelligence (AI) method, inspired by the approach presented in Mitchell *et al.* (2023), is introduced to identify various sound sources and predict one of the components in a circumplex appraisal of the sonic environment: annoyance. More specifically, this paper proposes a deep-learning model based on the cross-attention mechanism to simultaneously perform sound source classification (SSC) and annoyance rating prediction (ARP) for end-to-end inference of sound sources and annoyance rates in soundscapes. SSC has been widely used for audio event recognition (Kong *et al.*, 2020; Ren *et al.*, 2017) and acoustic scene classification (Barchiesi *et al.*, 2015; Hou *et al.*, 2022a; Mesaros *et al.*, 2018b). In this work, we will augment it with ARP, aiming to predict the overall appraisal of the soundscape along the annoyance axis.

In soundscapes with complex acoustic environments, source-related SSC and human-perception-related ARP are commonly used techniques for understanding how people perceive and respond to sounds in soundscapes. To accurately identify these various audio events, deep learning-based convolutional neural networks (Li *et al.*, 2019; Xu *et al.*, 2017), recurrent neural networks (Parascandolo *et al.*, 2016), convolutional recurrent neural networks (Li *et al.*, 2020), and Transformer (Vaswani *et al.*, 2017) with multi-head attention are used in SSC-related detection and classification of acoustic scenes and events (DCASE) challenges (Mesaros *et al.*, 2018a; Politis *et al.*, 2021). Recently, with the aid of large-scale audio datasets, e.g., AudioSet (Gemmeke *et al.*, 2017), and diverse audio pre-trained models [such as convolution-based PANNs (Kong *et al.*, 2020) and Transformer-based AST (Gong *et al.*, 2021)], deep learning-based approaches have made great improvement in SSC tasks. However, most of these SSC-related studies focus on recognizing sound sources without considering whether they are annoying to humans. This paper proposes a joint SSC and ARP approach, expanding SSC to include subjective human perception.

An intuitive observation is that in real-life soundscapes, loud sounds naturally attract more human attention than quieter sounds. For example, on the side of the street, the sound of roaring cars will capture people's attention more than the sound of small conversations on the corner. Therefore, this paper exploits the loudness-related root mean square value (RMS) (Mulimani and Koolagudi, 2018) and Mel spectrograms (Bala *et al.*, 2010) features, which conform to human hearing characteristics, to predict the objective sound sources and perceived annoyance ratings. The proposed model uses convolutional blocks to extract high-level representations of the two features and a cross-attention module to fuse their semantic representations. Based on the proposed model, this paper explores the following research questions (RQs):

(1) RQ1: Can the model's performance be improved using two acoustic features?
(2) RQ2: How does the performance of the proposed model compare with other models on the ARP task and the SSC task, as well as the joint ARP and SSC tasks? Does the cross-attention-based fusion module in the model work well?
(3) RQ3: Does the proposed model capture the relationships between sound sources and annoyance ratings? What are the relationships between sound sources, annoyance ratings, and sound levels?
(4) RQ4: How does the proposed model respond to adding unknown sounds to the soundscape?

The paper is organized as follows. Section II introduces the proposed method. Section III describes the baselines, dataset and training setup. Section IV analyzes and discusses the results with research questions. Section V draws conclusions.

## II. METHOD

This section introduces the proposed model DCNN-CaF: the dual-branch convolutional neural network (DCNN) with cross-attention-based fusion (CaF). First, we introduce how to extract audio representations from the input audio clips, and then perform CaF on audio representations. Finally, we use different loss functions to train task-dependent branches of the model to complete the classification-based SSC task and the regression-based ARP task.

### A. Audio representation extraction

Since the Mel spectrograms common in sound source-related tasks and RMS features that can reflect the energy of sound sources are used in this paper, there are two branches of inputs to the DCNN-CaF model to extract high-level representations of the two acoustic features separately, as shown in Fig. 1. Inspired by the excellent performance of pure convolution-based pretrained audio neural networks (PANNs) (Kong *et al.*, 2020) in audio-related tasks, a convolutional structure similar to that in PANNs is used in Fig. 1 to extract the representation of the input acoustic features. Specifically, the dual-input model in Fig. 1 uses 4-layer convolutional blocks. Each convolutional block contains two convolutional layers with global average pooling (GAP). The representations of Mel spectrograms and RMS features generated by the convolution block, $R_m$ and $R_r$, are fed to the attention-based fusion module to generate representations suitable for the ARP task. The embeddings of the sound source generated by the mapping of $R_m$ through the embedding layer will be input into the final sound source classification layer to complete the SSC task.

### B. Cross-attention-based fusion

The cross-attention fusion module in this paper is based on the multi-headed attention (MHA) in Transformer (Vaswani *et al.*, 2017). MHA allows models to jointly focus on representations at different positions in different subspaces. Following the description in Transformer (Vaswani

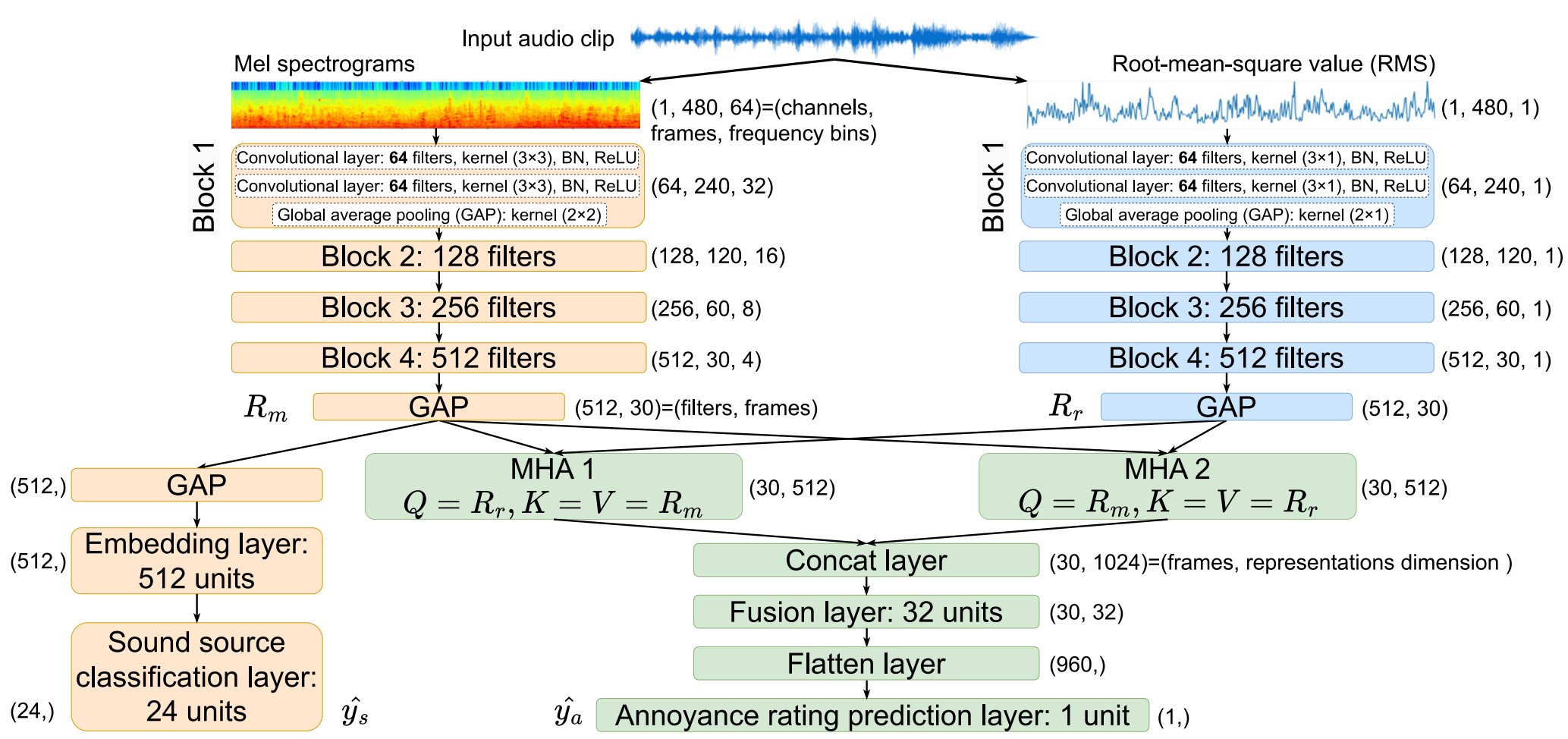

FIG. 1. (Color online) The proposed dual-branch convolutional neural network with cross-attention-based fusion (DCNN-CaF). The dimension of the output of each layer is shown.

et al., 2017), MHA is calculated on a set of queries (**Q**), keys (**K**), and values (**V**),

$$MHA(\mathbf{Q},\mathbf{K},\mathbf{V}) = Concat(head_1,\dots,head_h)\mathbf{w}^O, \quad (1)$$

where

$$head_i = A(\mathbf{Q}\mathbf{w}_i^Q,\mathbf{K}\mathbf{w}_i^K,\mathbf{V}\mathbf{w}_i^V), \quad (2)$$

$$A(\mathbf{Q}\mathbf{w}_i^Q,\mathbf{K}\mathbf{w}_i^K,\mathbf{V}\mathbf{w}_i^V) = \Phi\left(\mathbf{Q}\mathbf{w}_i^Q\mathbf{K}\mathbf{w}_i^{KT}/\sqrt{d}\right)\mathbf{V}\mathbf{w}_i^V, \quad (3)$$

where $head_i$ represents the output of the $i$th attention head for a total number of $h$ heads. $\mathbf{W}_i^Q$, $\mathbf{W}_i^K$, $\mathbf{W}_i^V$, and $\mathbf{W}^O$ are learnable weights. For MHA in the encoder, **Q**, **K**, and **V** come from the same place, at this point, the attention in MHA is called self-attention (Vaswani *et al.*, 2017). All the parameters (such as $h=8$, $d_k$, and $d_v = d_{model}/h = 512/8 = 64$, etc.) follow the default settings of Transformer (Vaswani *et al.*, 2017).

From the corresponding dimensions of the output of each layer in Fig. 1, it can be seen that the dimensions of $R_m$ and $R_r$ are both (512, 30), which correspond to the number of filters of the previous convolutional layer and the number of frames, respectively. After a series of convolutional layers operations, the input 480 frames of Mel spectrograms and RMS features are extracted into audio representations with a time length of 30 frames. This means that in MHA, the time step of each head $head_i$ is also 30. To obtain the representation of $R_m$ and $R_r$ based on the mutual attention of $R_m$ and $R_r$ collaboratively, in MHA1 in Fig. 1,

$$Q = R_m, \quad K = V = R_r. \quad (4)$$

In contrast, in MHA2,

$$Q = R_r, \quad K = V = R_m. \quad (5)$$

The cross-attention-adjusted representations of $R_m$ and $R_r$ are simply concatenated together and fed into the fusion layer to obtain higher acoustic representations containing the semantics of $R_m$ and $R_r$.

### C. The loss function of the DCNN-CaF model

The model proposed in this paper performs two tasks simultaneously, SSC and ARP. Given that the output of the sound source classification layer is $\hat{y}_s$, and its corresponding label is $y_s$, referring to the previous work (Hou and Botteldooren, 2022), the binary cross-entropy (BCE) is used as the loss function for the SSC task,

$$\mathcal{L}_{\text{SSC}} = BCE(\hat{y}_s, y_s). \quad (6)$$

Given the prediction output from the annoyance rating prediction layer is $\hat{y}_a$ and its corresponding label is $y_a$, the mean squared error (MSE) (Wallach and Goffinet, 1989) is used as a loss function for the ARP task to measure the distance between the predicted and the human-annotated annoyance ratings,

$$\mathcal{L}_{\text{ARP}} = MSE(\hat{y}_a, y_a). \quad (7)$$

Then, the final loss function of the DCNN-CaF model in this paper is

$$\mathcal{L} = \mathcal{L}_{\text{SSC}} + \mathcal{L}_{\text{ARP}}. \quad (8)$$

## III. DATASET, BASELINE, AND EXPERIMENTAL SETUP

### A. Dataset

To the best of our knowledge, DeLTA (Mitchell *et al.*, 2022) is the only publicly available dataset that includes both ground-truth sound source labels and human annoyance rating scores, so we use it in the paper. DeLTA comprises 2890 15-s binaural audio clips collected in urban public spaces across London, Venice, Granada, and Groningen. A remote listening experiment performed by 1221 participants was used to label the DeLTA recordings. In the listening experiment, participants listened to 10 15-s binaural recordings of urban environments, assessed whether they contained any of the 24 classes of sound sources, and then provided an annoyance rating (continuously from 1 to 10). Participants were given labels for 24 classes of sound sources, including: *Aircraft*, *Bells*, *Bird tweets*, *Bus*, *Car*, *Children*, *Construction*, *Dog bark*, *Footsteps*, *General traffic*, *Horn*, *Laughter*, *Motorcycle*, *Music*, *Non-identifiable*, *Rail*, *Rustling leaves*, *Screeching brakes*, *Shouting*, *Siren*, *Speech*, *Ventilation*, *Water*, and *Other*, adapted from the taxonomy developed by Salamon *et al.* (2014). In the listening experiment, each recording was evaluated by two to four participants, with an average of 3.1 recognized sound sources per recording. For more detailed information about DeLTA, please see (Mitchell *et al.*, 2022). During the training of models in this paper, the training, validation, and test sets contain 2081, 231, and 578 audio clips, respectively.

### B. Baseline for annoyance rating prediction (ARP) task

To compare the performance of the proposed deep-learning-based method with traditional approaches in soundscape-related studies, we employ five regression methods inspired by their performance in annoyance prediction in soundscape research (Al-Shargabi *et al.*, 2023; Iannace *et al.*, 2019; Morrison *et al.*, 2003; Szychowska *et al.*, 2018; Zhou *et al.*, 2018) to perform the ARP task based on A-weighted equivalent sound pressure levels. They are linear regression, support vector regression (SVR), decision tree (DT), k-nearest neighbours (KNN), and random forest. Linear regression is a fundamental and interpretable model that assumes a linear relationship between input features (in this case, sound levels) and the target variable (annoyance ratings). SVR is particularly effective when dealing with complex relationships between input features and target variables. Decision tree regression is known for its ability to handle non-linear relationships and interactions among features. Random forest regression is an ensemble method that combines multiple decision trees to improve predictive accuracy and reduce overfitting. KNN regression can work well when there is a relatively small dataset and in low-dimensional spaces.

### C. Baseline for sound source classification (SSC) task

In SSC-related research, deep learning convolutional neural network (CNN)-based models have achieved wide-spread success, and recently, Transformer-based models become dominant. Therefore, for the SSC task, the classical CNN-based YAMNet (Plakal and Ellis, 2023) and PANN (Kong *et al.*, 2020), and Transformer-based AST (Gong *et al.*, 2021) are used as baselines. Since YAMNet, PANN, and AST are trained on the large-scale AudioSet (Gemmeke *et al.*, 2017), the last layer of YAMNet, PANN, and AST has 527, 521, and 527 units for output, respectively. In contrast, the SSC task in this paper has only 24 classes of audio events, so we modify the number of units in the last layer of all three to 24, and then fine-tune the models on the DeLTA dataset.

### D. Baseline for joint ARP and SSC task

This paper first attempts to use the artificial intelligence (AI)-based model to simultaneously perform sound source classification and annoyance rating prediction. Therefore, this paper adopts deep neural networks (DNN), convolutional neural networks (CNN), and CNN-Transformer as baselines for comparison.

#### *1. Deep neural networks (DNN)*

The DNN consists of two branches. Each branch contains four fully connected layers and ReLU functions (Boob *et al.*, 2022), where the number of units in each layer is 64, 128, 256, and 512, respectively. The outputs of the final fully connected layer of the two branches are concatenated and combined to feed to the SSC and ARP layers, respectively.

#### *2. Convolutional neural networks (CNN)*

Similar to DNN, the compared CNN also consists of two branches. Each branch includes two convolutional layers, where the number of filters in each convolutional layer is 32 and 64, respectively. The outputs of the convolutional layers are concatenated and combined to feed to the SSC and ARP layers, respectively.

#### *3. CNN-Transformer*

The CNN-Transformer is based on CNN, and an Encoder from Transformer (Vaswani *et al.*, 2017) is added after the final convolutional layer in CNN. After the output of the Encoder is flattened, it is fed to the SSC and ARP layers, respectively.

### E. Training setup and metric

The 64-filter banks logarithmic Mel-scale spectrograms (Bala *et al.*, 2010) and frame-level root mean square values (RMS) (Mulimani and Koolagudi, 2018) are used as the acoustic features in this paper. A Hamming window length of 46 ms and a window overlap of 1/3 (Hou *et al.*, 2022a) are used for each frame. A batch size of 64 and Adam optimizer (Kingma and Ba, 2015) with a learning rate of 1e−3 are used to minimize the loss in the proposed model. The model is trained for 100 epochs.

The SSC is a classification task, so accuracy (*Acc*), *F-score*, and threshold-free area under curve (*AUC*) are used to evaluate the classification results. The ARP is viewed as a regression task in this paper, so mean absolute error (*MAE*) and root mean square error (*RMSE*) are used to measure the regression results. Higher *Acc*, *F-score*, *AUC* and lower *RMSE*, *MAE* indicate better performance. Models and more details are available on the project webpage (Hou, 2023).

## IV. RESULTS AND ANALYSIS

This section analyzes the performance of the proposed method based on the following research questions.

### A. Can the model's performance be improved using two acoustic features?

Two kinds of acoustic features are used in this paper, the Mel spectrograms that approximate the characteristics of human hearing and the RMS features that characterise the acoustic level. Table I shows the ablation experiments of the two acoustic features on the proposed DCNN-CaF model to specifically present the performance of the DCNN-CaF model based on different features. When only a single feature is used, the input of the DCNN-CaF model is the corresponding single branch.

As shown in Table I, the DCNN-CaF model performs the worst on the ARP and SSC tasks when using only the 46 ms interval RMS features, which are related to instantaneous loudness. This is apparently caused by the lack of spectral information, which is embedded in the Mel spectrograms and omitted from the RMS features. The dimension of the frame-level RMS used in this paper is (T, 1), where T is the number of frames. Compared with Mel spectrograms with a dimension of (T, 64), the spectral information contained in the loudness-related one-dimensional RMS features is also scarcer. This factor makes it difficult for the model to distinguish the 24 types of sound sources and predict annoyance from real-life different sound sources in the DeLTA dataset based only on the RMS features alone. The DCNN-CaF using Mel spectrograms outperforms the results of its corresponding RMS features overall. While DCNN-CaF combining Mel spectrograms and RMS features achieves the best results, which clarifies that using these two acoustic features benefits the model's performance on SSC and ARP tasks. Thus, adding energy level-related information to the sound recognition improves annoyance prediction as expected, but it also slightly improves sound source recognition.

TABLE I. Ablation study on the acoustic features.

| | Acoustic feature | | ARP | | SSC | | |
|---|---|---|---|---|---|---|---|
| # | Mel | RMS | *MAE* | *RMSE* | *AUC* | *F-score* (%) | *Acc* (%) |
| 1 | ✓ | ✗ | 1.00 ± 0.15 | 1.18 ± 0.12 | 0.89 ± 0.01 | 61.12 ± 5.45 | 90.98 ± 0.99 |
| 2 | ✗ | ✓ | 1.08 ± 0.14 | 1.27 ± 0.10 | 0.79 ± 0.02 | 53.64 ± 3.20 | 89.33 ± 1.73 |
| 3 | ✓ | ✓ | **0.84** ± 0.12 | **1.05** ± 0.13 | **0.90** ± 0.01 | **67.20** ± 3.16 | **92.52** ± 0.87 |

TABLE II. Performance of models trained only for the SSC task.

| Model | YAMNet | PANN | AST | The proposed DCNN-CaF |
|---|---|---|---|---|
| Parameters (Million) | 3.231 | 79.723 | 86.207 | 4.961 (Only SSC branch) |
| AUC | 0.87 ± 0.02 | 0.90 ± 0.02 | 0.85 ± 0.3 | **0.92** ± 0.01 |
| F-score (%) | 63.95 ± 2.67 | 66.52 ± 2.31 | 56.91 ± 2.84 | **67.08** ± 2.23 |
| Acc (%) | 90.56 ± 1.96 | 91.69 ± 1.44 | 89.39 ± 1.92 | **92.34** ± 1.58 |

### B. How does the performance of the proposed model compare with other models on the ARP task and the SSC task, as well as the joint ARP and SSC tasks? Does the cross-attention-based fusion module in the model work well?

Table II presents the results of classical pure convolution-based YAMNet and PANN (Kong *et al.*, 2020), and Transformer-based AST (Gong *et al.*, 2021), on the SSC task. YAMNet, PANN, and AST are trained based on Mel spectrograms. For a fair comparison, the proposed DCNN-CaF only uses the left SSC branch of the input Mel spectrograms.

In Table II, both YAMNet and DCNN-CaF are lightweight models compared to PANN and AST. Relative to Transformer-based AST, the number of parameters of DCNN-CaF is reduced by $(86.207-4.961)/86.207\times 100\% \approx 94\%$. Compared to YAMNet, PANN, and AST, which have deeper layers than DCNN-CaF, the shallow DCNN-CaF achieves better results on the SSC task, which may be due to the relatively small dataset used in this paper, and large and deep models are prone to overfitting during the training process.

Table III shows the joint ARP and SSC baselines proposed in Sec. III D. For a fairer comparison, the DNN, CNN, and CNN-Transformer in Sec. III D also use a dual-input branch structure to simultaneously use the two acoustic features of Mel spectrograms and RMS to complete the SSC and ARP tasks.

As shown in Table III, the CNN based on the convolutional structure outperforms the DNN based on multi-layer perceptrons (MLP) (Kruse *et al.*, 2022) on both tasks, which reflects that the convolutional structure is more effective than the MLP structure in extracting acoustic representations. While performing well on the ARP regression task, the CNN-Transformer combining convolution and an Encoder from Transformer has the worst result corresponding to the SCC task for real-life 24-class sound source recognition. This may be because the DeLTA dataset used in this paper is not large enough to allow the Transformer Encoder with MHA (Vaswani *et al.*, 2017) to play its expected role. Previous work has also shown that Transformer-based models tend to perform less well on small datasets (Hou *et al.*, 2022b). Finally, compared to these common baseline models, DCNN-CaF achieves better results on both SSC and ARP.

Next, we explore the performance of traditional A-weighted equivalent sound pressure level ($L_{Aeq}$)-based methods for annoyance prediction (ARP task). Thus, we extract the sound levels of audio clips in the DeLTA dataset and utilize them as features to predict annoyance ratings, as shown in Table IV. Note that sound level is a clip-level feature, while the proposed DCNN-CaF only accepts frame-level features as input. Therefore, the proposed DCNN-CaF, which cannot input coarse-grained clip-level $L_{Aeq}$-based sound level features, is omitted in Table IV.

Compared to the other models in Table IV, the support vector regression (SVR) achieves the best performance on the ARP task. This may be attributed to its robustness in handling outliers and its ability to effectively model nonlinear relationships (Izonin *et al.*, 2021). In summary, the $L_{Aeq}$-based traditional approaches in Table IV show competitive performance on the ARP task, and their performance is close to the deep learning neural network-based methods in Table III.

TABLE III. Comparison of different models for joint SSC and ARP tasks on the DeLTA dataset.

| Model | Param. (M) | ARP | | SSC | | |
|---|---|---|---|---|---|---|
| | | *MAE* | *RMSE* | *AUC* | *F-score (%)* | *Acc (%)* |
| DNN | 0.376 | 1.06 ± 0.19 | 1.32 ± 0.08 | 0.87 ± 0.01 | 49.57 ± 8.78 | 89.30 ± 2.21 |
| CNN | 1.271 | 0.94 ± 0.06 | 1.15 ± 0.06 | 0.88 ± 0.01 | 52.03 ± 3.09 | 90.34 ± 0.46 |
| CNN-Transformer | 17.971 | 1.00 ± 0.13 | 1.14 ± 0.09 | 0.86 ± 0.02 | 46.89 ± 6.32 | 88.56 ± 1.24 |
| DCNN-CaF | 7.614 | **0.84** ± 0.12 | **1.05** ± 0.13 | **0.90** ± 0.01 | **67.20** ± 3.16 | **92.52** ± 0.87 |

TABLE IV. Comparison of sound level-based approaches on the ARP task.

| Method | Linear regression | Decision tree | SVR | Random forest | KNN |
|---|---|---|---|---|---|
| MAE | 1.03 | 1.45 | 1.01 | 1.27 | 1.13 |
| RMSE | 1.26 | 1.84 | 1.26 | 1.60 | 1.41 |

To intuitively present the results of DCNN-CaF for the annoyance rating prediction, Fig. 2 visualizes the gap between the annoyance ratings predicted by DCNN-CaF and the corresponding ground-truth annoyance ratings. The red point representing the predicted value and the blue point indicating the true label in Fig. 2 mostly match well, indicating that the proposed model successfully regresses the annoyance ratings in the real-life soundscape.

For an in-depth analysis of the performance of the DCNN-CaF, Fig. 3 further visualizes the attention distribution from the cross-attention-based fusion module on some test samples. As described in Sec. III B, the number of time steps of each head in the multi-head attention (MHA) is 30, which comes from 30 frames in the dimensions of the representations of Mel spectrograms and RMS features before MHA. Therefore, in DCNN-CaF, the dimension of the attention matrix of each head in MHA is $30 \times 30$. Figure 3 visualizes the distribution of attention in the same head number from MHA1 and MHA2. From the distribution of attention in the subgraphs of Fig. 3, it can be seen that the MHA1, which uses $R_r$ to adjust $R_m$ and MHA2, which uses $R_m$ to adjust $R_r$, complement each other. For example, for sample #1 in Fig. 3, the attention of MHA1 in subfigure (1) is mainly distributed on the left side, while the attention of MHA2 in subfigure (2) is predominantly distributed on the right side. For the same sample, MHA1 and MHA2 with different attention perspectives match each other well. The results in Fig. 3 illustrate that the proposed DCNN-CaF model successfully pays different attention to the information of different locations of two kinds of acoustic features based on the cross-attention module, which is beneficial for the fusion of these acoustic features. For more visualizations of all the attention distributions of the 8 heads of MHA1 and MHA2, please see the *project webpage*.

### C. Does the proposed model capture the relationships between sound sources and annoyance ratings? What are the relationships between sound sources, annoyance ratings, and sound levels?

To identify which of the 24 classes of sounds is most likely to cause annoyance, we first analyze the relationship between the sound identified by the model and the annoyance it predicts. Then, the predictions from the model are compared to the human classification. Specifically, we first use Spearman's rho (David and Mallows, 1961) to analyze the correlation between the probability of various sound sources predicted by the model and the corresponding annoyance ratings. Then, we calculate the distribution of sound sources at different annoyance ratings, and further verify the model's predictions based on human-annotated sound sources and annoyance rating labels.

*Correlation analysis between the model's sound source classification and annoyance ratings*. A Shapiro-Wilk (with $\alpha = 0.05$) statistic test (Hanusz *et al.*, 2016) is performed before a correlation analysis of the model's predicted sound sources and annoyance ratings on the test set. The results of the Shapiro-Wilk statistic test showed no evidence that the model's predictions conform to a normal distribution. Therefore, a non-parametric method named Spearman's rho (David and Mallows, 1961) is used for correlation analysis. The Spearman's rho correlation analysis in Table V shows that the recognition of some sounds is significantly correlated with the predicted annoyance rating. Specifically, the presence of sound sources such as *Children*, *Water*, *Rail*, *Construction*, *Siren*, *Shouting*, *Bells*, *Motorcycle*, *Music*, *Car*, *General traffic*, *Screeching brakes*, *Horn*, and *Bus* is positively correlated with the annoyance rating. The presence of sound sources such as *Ventilation*, *Footsteps*, *Dog bark*, *Bird tweet*, *Rustling leave*, *Non-identifiable*, and *Other* is negatively correlated with the annoyance rating. As for sound sources such as *Speech*, *Aircraft*, and *Laughter*, there is no significant correlation between them being present and annoyance rating. Further correlation analysis indicates that the sound source *Bus* shows the highest positive correlation with the annoyance rating, with a correlation coefficient of 0.712. In contrast, the sound source *Rustling leave* shows

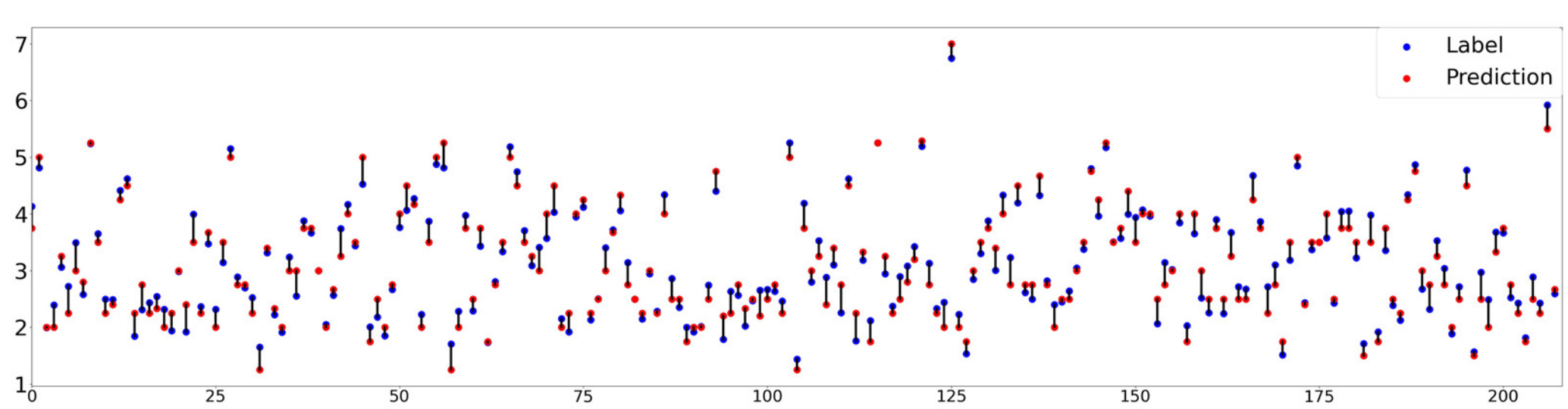

FIG. 2. (Color online) Scatter plot of annoyance ratings of model predictions and human-annotated labels on some samples in the test set. Black vertical lines indicate gaps between two points.

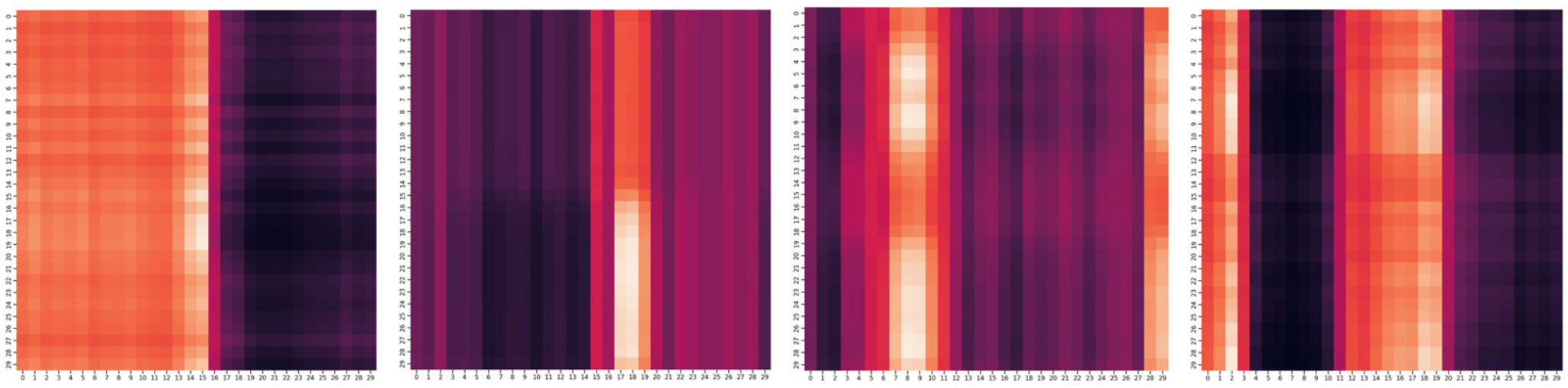
(1) Sample #1: MHA 1 (2) Sample #1: MHA 2 (3) Sample #2: MHA 1 (4) Sample #2: MHA 2

FIG. 3. (Color online) Attention distributions of the cross-attention-based fusion module in DCNN-CaF on audio clips from the test set. The subgraphs (1, 2) and (3, 4) are from (MHA1, MHA2) of sample #1 and #2, respectively. Brighter colours represent larger values.

the highest negative correlation with the annoyance rating, with a coefficient of –0.731.

*Verifying the model's predictions based on human-perceived manually annotated labels*. Based on the correlation analysis of the model's predictions, some sound sources are more likely to cause people annoyance than others. To investigate the consistency of the correlation analysis results between the model-based and the human-annotated labels-based, we calculate the distribution of sound sources at different annoyance rating levels based on the human-annotated labels to explore the correlations between the sound source and the annoyance levels, as shown in Fig. 4.

Given that the mean value of the annoyance rating by humans on the test set is $\mu$, for the $i$th class of sound source $s_i$, the total number of occurrences in audio samples with an annoyance rating less than or equal to $\mu$ is $n_{i,l}$, and the total number of occurrences in audio samples with an annoyance rating greater than $\mu$ is $n_{i,h}$. $N_i = n_{i,l} + n_{i,h}$, $N_i$ is the total number of samples containing the sound source $s_i$. Then, the probability of the sound source occurring in the samples where annoyance is lower than or equal to $\mu$ is

$$P(x \le \mu | s_i) = \frac{n_{i,l}}{n_{i,l} + n_{i,h}} = \frac{n_{i,l}}{N_i}, \tag{9}$$

where $x$ represents the annoyance rating for fragments containing the sound source, $s_i$. Correspondingly, the probability of it occurring in samples higher than $\mu$ is

$$P(x > \mu | s_i) = \frac{n_{i,h}}{N_i} = 1 - P(x \le \mu | s_i). \tag{10}$$

TABLE V. Spearman's rho correlation coefficients on DeLTA.

| Sound source | Model predicted annoyance Correlation (r) | Annotated by human $P(x \le \mu)$ | $P(x > \mu)$ | Samples (N) | Sound levels Correlation(r) |
|---|---|---|---|---|---|
| Children | 0.12[a] | 0.35 | 0.65 | 71 | 0.01 |
| Water | 0.18[a] | 0.39 | 0.61 | 69 | 0.02 |
| Rail | 0.26[a] | 0.33 | 0.67 | 15 | 0.03 |
| Construction | 0.27[a] | 0.43 | 0.57 | 54 | 0.01 |
| Siren | 0.27[a] | 0.57 | 0.43 | 21 | 0.02 |
| Shouting | 0.35[a] | 0.20 | 0.80 | 35 | 0.01 |
| Bells | 0.33[a] | 0.37 | 0.63 | 19 | 0.01 |
| Motorcycle | 0.47[a] | 0.24 | 0.76 | 37 | 0.05 |
| Music | 0.50[a] | 0.21 | 0.79 | 43 | 0.04 |
| Car | 0.52[a] | 0.35 | 0.65 | 79 | 0.04 |
| General traffic | 0.58[a] | 0.38 | 0.62 | 240 | 0.02 |
| Screeching brakes | 0.63[a] | 0.25 | 0.75 | 8 | 0.05 |
| Horn | 0.70[a] | 0.28 | 0.72 | 21 | 0.06 |
| Bus | **0.71**[a] | 0.11 | **0.89** | 19 | 0.04 |
| Ventilation | −0.20[a] | 0.50 | 0.50 | 34 | 0.01 |
| Footsteps | −0.42[a] | 0.63 | 0.37 | 223 | −0.03 |
| Non-identifiable | −0.43[a] | 0.65 | 0.35 | 52 | −0.01 |
| Other | −0.49[a] | 0.59 | 0.41 | 85 | −0.04 |
| Dog bark | −0.56[a] | 0.75 | 0.25 | 8 | 0.00 |
| Bird tweet | −0.69[a] | 0.74 | 0.26 | 160 | −0.01 |
| Rustling leaves | **−0.73**[a] | **0.83** | 0.17 | 18 | −0.06 |
| Speech | 0.05 | 0.46 | 0.54 | 368 | −0.02 |
| Aircraft | −0.02 | 0.45 | 0.55 | 20 | 0.03 |
| Laughter | −0.05 | 0.42 | 0.58 | 78 | −0.02 |

[a]Statistical significance at the 0.01 level.

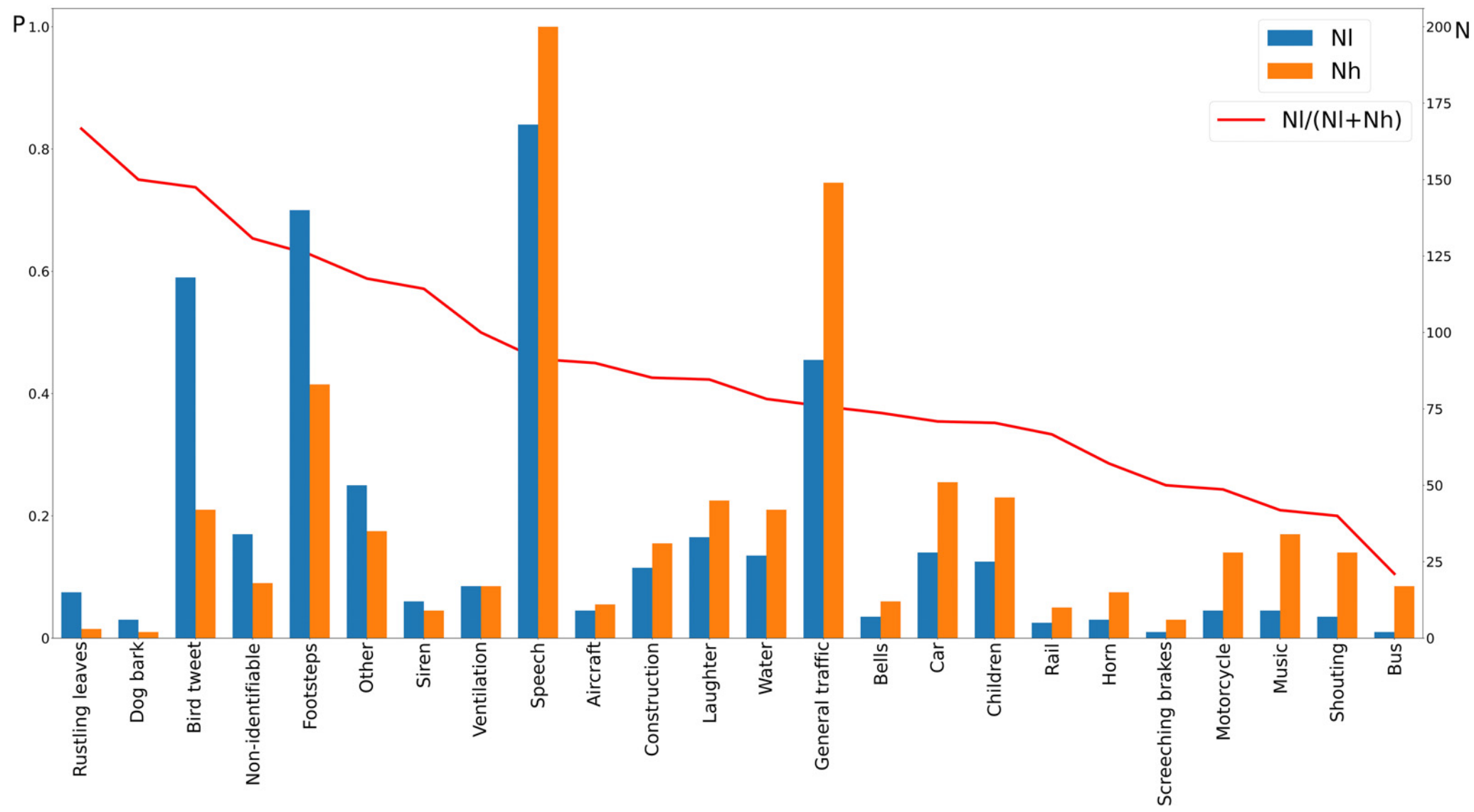

FIG. 4. (Color online) Distribution of the number of samples from different sources at low and high annoyance rating levels, that is, $n_{i,l}$ and $n_{i,h}$ (shown as Nl and Nh), and the corresponding $P(x \leq \mu|s_i)$ curves.

Table V comprehensively shows the probability distribution of 24 classes of sound sources at different levels of annoyance rating according to human perception.

The probability distribution of sound sources at different annoyance rating levels in Table V reveals that according to people's real feelings, *Rustling leaves* sounds have the highest probability in the low annoyance rating level ($x \leq \mu$), while *Bus* sounds have the highest probability in the high annoyance rating level ($x > \mu$). This successfully verifies the correctness of the above model-based correlation analysis between sound sources and annoyance ratings. Furthermore, Table V also shows that *Children*, *Water*, *Rail*, *Construction*, *Shouting*, *Bells*, *Motorcycle*, *Music*, *Car*, *General traffic*, *Screeching brakes*, *Horn*, and *Bus* are more likely to occur in the high annoyance rating level, while *Footsteps*, *Dog bark*, *Bird tweet*, *Rustling leave*, *Non-identifiable*, and *Other* are more likely to occur in the low annoyance rate level. *Speech*, *Aircraft*, and *Ventilation* have a similar probability of occurring in the high and low annoyance levels, implying that they may be more prevalent in the soundscape of the test set. In short, both the proposed model-based and human-perceived-based analyses showed similar trends regarding which sound sources are most strongly associated with annoyance levels. The consistency between the two analyses in identifying sound sources most strongly associated with annoyance ratings indicates that the proposed model performs well in predicting the relationships between sound sources and annoyance ratings.

*Correlations between sound level and annoyance rating*. In addition to exploring the correlation between sound sources and annoyance ratings, we further analyze the correlation between fragment-level A-weighted equivalent sound pressure level ($L_{Aeq}$) and human-perceived annoyance based on Kendall's Tau (often referred to as Kendall's Tau rank correlation coefficient). The corresponding result is ($tau = 0.42, p < 0.001$). That is, there is a significant correlation between sound level and annoyance rating in the DeLTA dataset. This is not unexpected as the ARP baseline models based on $L_{Aeq}$ only in Table IV have some predictive power.

Next, we delve into the relationship between the probability of the presence of sound sources predicted by the model and sound levels. Table V shows that there is no significant Pearson correlation between sound sources and sound levels in the DeLTA dataset. That is, the 24 different classes of sound sources in the DeLTA dataset cannot be identified solely by relying on fragment-level sound level information.

*Case study*. Notably, there is a significant positive correlation between *Music* and annoyance ratings in Table V. According to the statistical results in DeLTA (Mitchell *et al.*, 2022), the average annoyance score for clips with *Music* sources is 4.01, while the average annoyance score for clips without *Music* sources is 3.29, which implies that most of the presence of *Music* in DeLTA causes an increase in annoyance rather than being relaxing. Previous studies also show that there are various types of annoying music in daily life (Trotta, 2020).

In order to analyze it in depth, we further filter out all audio clips containing music in DeLTA, totaling 222 15-s clips, with an average sound level of 81.8 dBA. We then analyze the relationship between the sound levels and annoyance ratings of these 222 music clips. The results show that under the condition of music source, there is a significant positive correlation between sound level and human-perceived annoyance ($tau = 0.18, p < 0.001$) in the DeLTA dataset. In summary, even though the *Music* is not significantly correlated with the sound level, it is weakly positively correlated with the sound level in Table V. In addition, the overall sound level in the audio clips where the *Music* source exists is high, and the sound level is significantly related to annoyance, which may contribute to the significant positive correlation between *Music* and annoyance presented in Table V.

In addition to sound level, characteristics of music, such as its style or genre, give people different listening experiences. Previous research reveals the role of music in inciting or mitigating antisocial behaviour, and that certain music genres can soothe or agitate individuals. Additionally, the perception of annoyance may also be affected, depending on the choice of music. For example, genres featuring heavy rhythms, known for their potential to evoke angry emotions (Areni, 2003; Cowen *et al.*, 2020), are often not favoured by listeners in an urban environment context. As highlighted in the study (Landström *et al.*, 1995), a key contributor to annoyance is the presence of a tonal component in the noise. Individuals exposed to intruding noise containing tonal elements tend to report higher levels of annoyance than those exposed to non-tonal noise. Furthermore, reported levels of annoyance tend to increase when the noise contains multiple tonal components. This observation suggests that tonal characteristics present in the sound source (and possibly also in the music) may be a contributing factor to the positive correlation between music and annoyance ratings in Table V.

### D. How does the proposed model respond to adding unknown sounds to the soundscape?

To investigate the generalization performance of the proposed DCNN-CaF, we randomly add 20 classes of sound sources as noise to the test set in this paper to explore the model's performance in predicting annoyance ratings in soundscapes with added unknown sound sources. To add a variety of sound source samples to the 578 audio clips in the test set, we first use 20 sound sources from the public ESC-50 dataset (Piczak, 2015) as additional noise sources, each source containing 40 5-s audio samples. Then, we randomly add the 5-s noise source samples to the 15-s audio files in the test set, and each 15-s audio file is randomly assigned 1 to 3 5-s audio samples from the same noise source. During the synthesis process, the signal-to-noise ratio (SNR) defaults to 0. In this way, we get 20 test sets containing different types of noise sources. Therefore, the total number of audio clips containing model-unknown noise is $20 \times 578 = 11\,560$, and the corresponding audio duration is about 48.2 h ($15\,\text{s} \times 11\,560 = 173\,400\,\text{s}$).

Figure 5 shows the average human-annotated annoyance rating for sounds in the test set, the average annoyance rating predicted by the model for the test set without external noise added, and the average annoyance rating predicted by the model for the test set added with 20 classes of noise. As shown in Fig. 5, without additional noise, the average annoyance rating of the 578 15-s audio clips in the test set in the soundscape predicted by the model is similar to that of human-perceived annoyance ratings. The standard deviation of our model prediction (the yellow line on the bars in Fig. 5) is smaller than the corresponding human-perceived annoyance, which intuitively demonstrates that the proposed model achieves a similar effect on the test set as the annoyance ratings from humans perception.

Adding the 20 types of sources at an SNR of 0, increases the sound level (i.e., RMS) and, therefore, would most probably increase the annoyance rating. If the model purely relied on the RMS value, as some other noise annoyance models do, it would predict the same increase for all sources. However, different annoyance levels are predicted depending on the source added, which intuitively corresponds better to human perception. The subtle sounds, such as *Pouring water* and *Keyboard typing*, are less likely to increase much annoyance. Compared to sound sources that are less likely to introduce human annoyance, such as *Water drops* and *Clapping*, the results in Fig. 5 illustrate that sound sources related to machines or engines increase annoyance ratings more strongly for the same increase in average sound level. Overall, the model's predictions in Fig. 5 are consistent with what can be expected, but its validity is not confirmed by experiments with human participation.

Figure 5 presents the performance of the DCNN-CaF model under artificially added unknown noise sources. However, the synthetic dataset in Fig. 5 is difficult to compare with real-life audio in terms of the realism and naturalness of the sound. To compare the performance of the proposed model on unknown data and, in particular, investigate its performance for predicting positive effects of soundscape augmentation, we test the DCNN-CaF on a real-world acoustic dataset from a road traffic noise environment (denoted as RTNoise) (Coensel *et al.*, 2011). The experiment in Coensel *et al.* (2011) shows that adding bird song and fountain sound can reduce human-perceived traffic noise loudness and increase perceived pleasantness. RTNoise contains recordings of freeways, major roads, and minor roads sounds and mixtures of these sounds with two bird choruses and two fountain sounds.

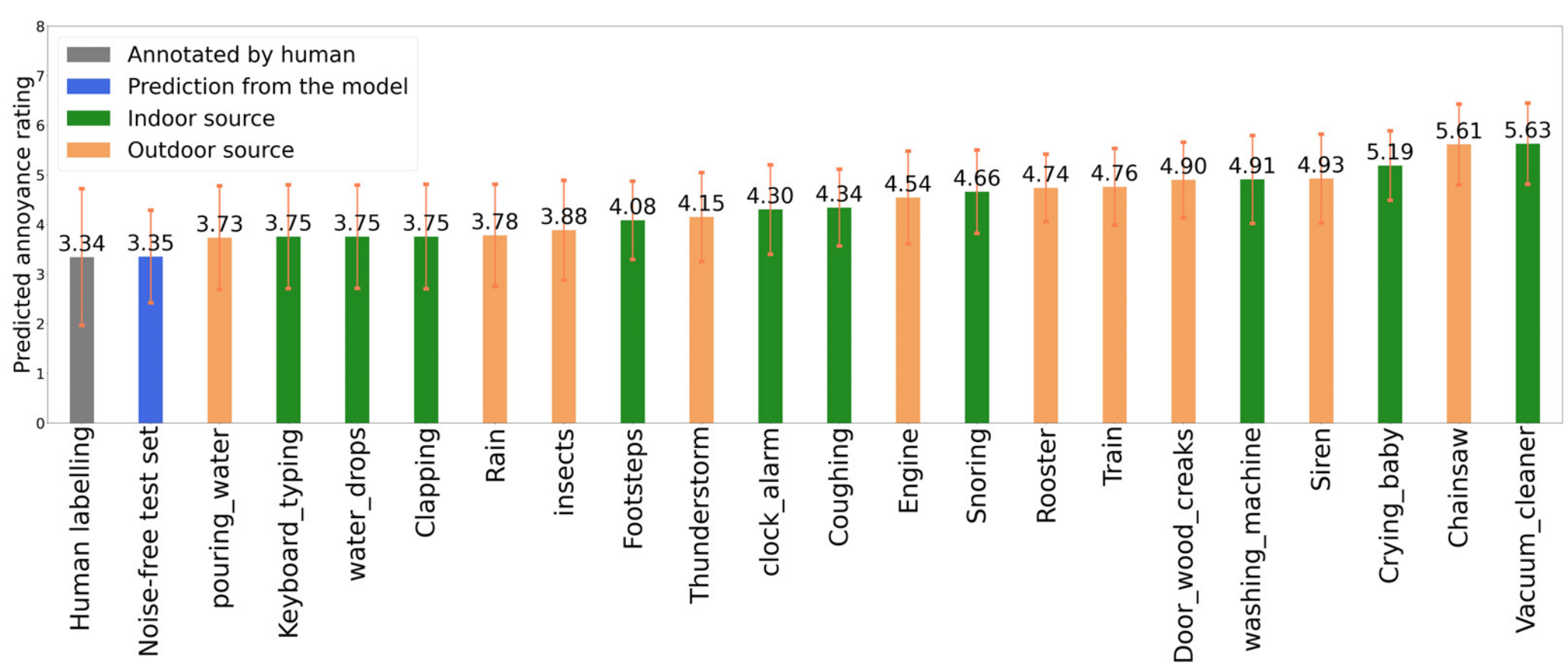

FIG. 5. (Color online) The mean and standard deviation of the predicted annoyance rating under different model-unknown noise sources.

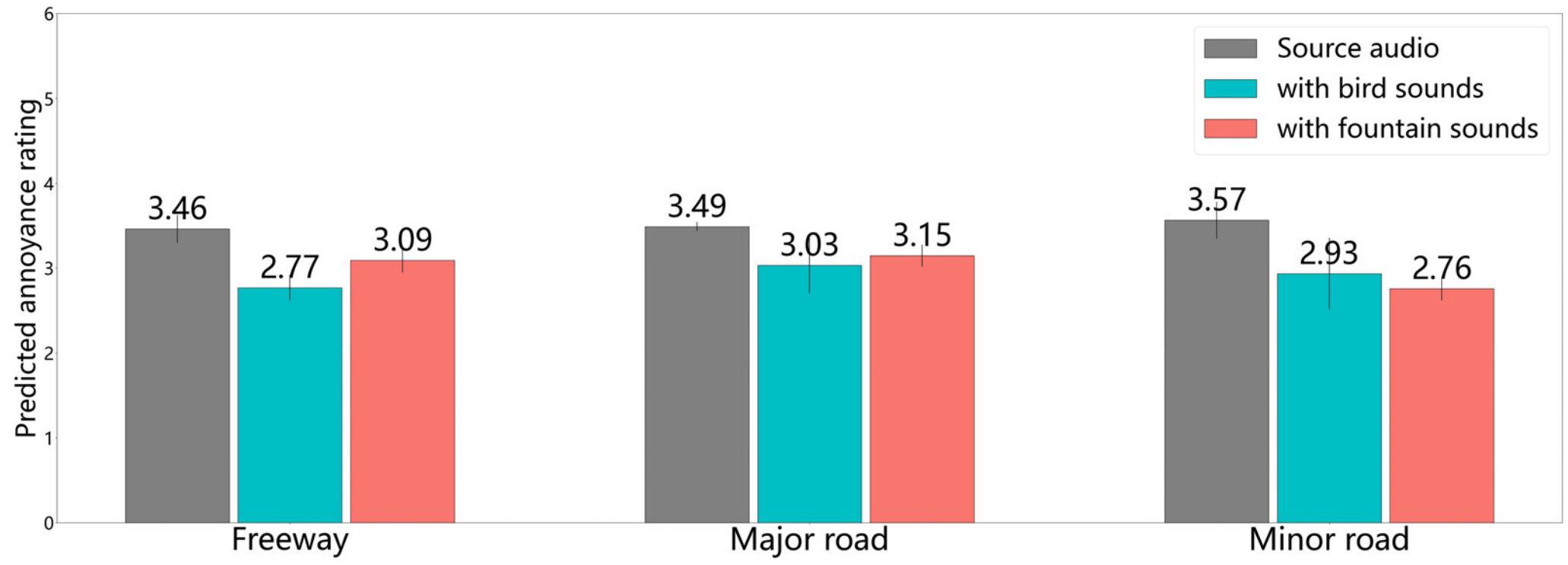

FIG. 6. (Color online) The performance of the proposed DCNN-CaF on real-life audio recordings.

As shown in Fig. 6, whether it is on the freeways, major roads, or minor roads, compared to the source audio clips, the predicted annoyance ratings of the audio clips with added bird sounds or fountain sounds will be reduced to varying degrees. Using the same sounds, tests with human listeners in Coensel *et al.* (2011) show a similar tendency for perceived traffic noise loudness for freeway sound, but this effect was not so prominent for major roads and minor roads. Human-rated pleasantness in the same experiment shows an opposite trend to the predicted trend in annoyance, which can be seen as the opposite. However, this prior experimental work also showed lower perceived loudness and higher pleasantness for the minor and major roads compared to freeway sound. The DCNN-CaF model does not seem to be able to distinguish between these types of sound. This could be caused by the poor control of noise levels in the online playback during the DeLTA data collection or by the shortness of the audio fragments (15 s) that do not allow the model to learn the difference between short car passages and continuous traffic. Other studies show that natural sounds, especially birdsong, can relax people (Van Renterghem, 2019); and under similar noise exposure conditions, respondents in neighbourhoods with more bird songs and fountains reported lower levels of annoyance (Qu *et al.*, 2023). The response of the proposed model in Fig. 6 to the sounds of birdsong and fountains in a real soundscape successfully matches this existing research.

## V. CONCLUSION

Soundscape characterization involves identifying sound sources and assessing human-perceived emotional qualities along multiple dimensions. This can be a time-consuming and expensive process when relying solely on human listening tests and questionnaires. This paper investigates the feasibility of using artificial intelligence (AI) to perform soundscape characterization without human questionnaires. Predictive soundscape models based on measurable features, such as the model proposed here, can enable perception-focused soundscape assessment and design in an automated and distributed manner, beyond what traditional soundscape methods can achieve. This paper proposes the cross-attention-based DCNN-CaF using two kinds of acoustic features to ensure the accuracy and reliability of the AI model, and simultaneously perform both the sound source classification task and the annoyance rating prediction task. The proposed AI model in this paper is trained on the DeLTA dataset, which contains sound source labels and human-annotated labels along one of the emotional dimensions of perception: annoyance.

Our experimental analysis demonstrates the following findings: (1) the proposed DCNN-CaF with dual-input branches using Mel spectrograms and loudness-related RMS features outperforms models using only one of these features. (2) On the sound source classification and annoyance rating prediction tasks, the DCNN-CaF with the attention-based fusion of two features outperforms DNN, CNN, and CNN-Transformer, which concatenate two features directly. In addition, attention visualization in the DCNN-CaF model shows that the cross-attention-based fusion module successfully pays different attention to the information of different acoustic features, which is beneficial for the fusion of these acoustic features. (3) Correlation analysis shows that the model successfully predicts the relationships between various sound sources and annoyance ratings, and these predicted relationships are consistent with those perceived by humans in the soundscape. (4) Generalization tests show that the model's ARP in the presence of model-unknown sources is consistent with expert expectations and can explain previous findings from the literature on soundscape augmentation. Future work involves extending the soundscape appraisal with other dimensions and taking into account more practical factors, such as participants' hearing and cultural and linguistic differences, to expand the training dataset to cover more acoustic scenarios. Furthermore, to improve the interpretability of the proposed model, the following work will try to visualize the learned weights of the model through heatmap analysis to clarify which neurons play a more critical role in learning to help explain the decision-making process of the model.

## ACKNOWLEDGMENTS

The WAVES Research Group received funding from the Flemish Government under the "Onderzoeksprogramma Artificiële Intelligentie (AI) Vlaanderen" programme.

Abraham, A., Sommerhalder, K., and Abel, T. (**2010**). "Landscape and well-being: A scoping study on the health-promoting impact of outdoor environments," Int. J. Public Health **55**, 59–69.

Acun, V., and Yilmazer, S. (**2018**). "Understanding the indoor soundscape of study areas in terms of users' satisfaction, coping methods and perceptual dimensions," Noise Cont. Eng. J. **66**(1), 66–75.

Al-Shargabi, A. A., Almhafdy, A., AlSaleem, S. S., Berardi, U., and Ali, A. A. M. (**2023**). "Optimizing regression models for predicting noise pollution caused by road traffic," Sustainability **15**(13), 10020.

Areni, C. S. (**2003**). "Examining managers' theories of how atmospheric music affects perception, behaviour and financial performance," J. Retail. Consumer Serv. **10**(5), 263–274.

Axelsson, Ö., Nilsson, M. E., and Berglund, B. (**2010**). "A principal components model of soundscape perception," J. Acoust. Soc. Am. **128**(5), 2836–2846.

Bala, A., Kumar, A., and Birla, N. (**2010**). "Voice command recognition system based on MFCC and DTW," Int. J. Eng. Sci. Technol. **2**(12), 7335–7342.

Barchiesi, D., Giannoulis, D., Stowell, D., and Plumbley, M. D. (**2015**). "Acoustic scene classification: Classifying environments from the sounds they produce," IEEE Signal Process. Mag. **32**(3), 16–34.

Beyer, A., Kamin, S. T., and Lang, F. R. (**2017**). "Housing in old age: Dynamical interactions between neighborhood attachment, neighbor annoyance, and residential satisfaction," J. Housing Elderly **31**(4), 382–393.

Boes, M., Filipan, K., De Coensel, B., and Botteldooren, D. (**2018**). "Machine listening for park soundscape quality assessment," Acta Acust. united Acust. **104**(1), 121–130.

Boob, D., Dey, S. S., and Lan, G. (**2022**). "Complexity of training ReLU neural network," Discrete Optim. **44**, 100620.

Brambilla, G., and Maffei, L. (**2010**). "Perspective of the soundscape approach as a tool for urban space design," Noise Control Eng. J. **58**(5), 532–539.

Bruce, N. S., and Davies, W. J. (**2014**). "The effects of expectation on the perception of soundscapes," Appl. Acoust. **85**, 1–11.

Carlsson, F., Karlson, B., Ørbaek, P., Österberg, K., and Östergren, P. (**2005**). "Prevalence of annoyance attributed to electrical equipment and smells in a Swedish population, and relationship with subjective health and daily functioning," Public Health **119**(7), 568–577.

Coensel, B. D., Vanwetswinkel, S., and Botteldooren, D. (**2011**). "Effects of natural sounds on the perception of road traffic noise," J. Acoust. Soc. Am. **129**(4), EL148–EL153.

Cowen, A. S., Fang, X., Sauter, D., and Keltner, D. (**2020**). "What music makes us feel: At least 13 dimensions organize subjective experiences associated with music across different cultures," Proc. Natl. Acad. Sci. U.S.A. **117**(4), 1924–1934.

Das, C. P., Swain, B. K., Goswami, S., and Das, M. (**2021**). "Prediction of traffic noise induced annoyance: A two-staged SEM-artificial neural network approach," Transp. Res. Part D: Transp. Environ. **100**, 103055.

David, F. N., and Mallows, C. L. (**1961**). "The variance of spearman's rho in normal samples," Biometrika **48**(1/2), 19–28.

Eek, F., Karlson, B., Österberg, K., and Östergren, P. (**2010**). "Factors associated with prospective development of environmental annoyance," J. Psychosom. Res. **69**(1), 9–15.

Fan, J., Thorogood, M., and Pasquier, P. (**2017**). "Emo-soundscapes: A dataset for soundscape emotion recognition," in *International Conference on Affective Computing and Intelligent Interaction*, pp. 196–201.

Fan, J., Thorogood, M., Riecke, B. E., and Pasquier, P. (**2015**). "Automatic recognition of eventfulness and pleasantness of soundscape," in *Proceedings of the Audio Mostly 2015 on Interaction with Sound*, pp. 1–6.

Fan, J., Tung, F., Li, W., and Pasquier, P. (**2018**). "Soundscape emotion recognition via deep learning," in *Proceedings of the Sound and Music Computing*.

Fang, X., Gao, T., Hedblom, M., Xu, N., Xiang, Y., Hu, M., Chen, Y., and Qiu, L. (**2021**). "Soundscape perceptions and preferences for different groups of users in urban recreational forest parks," Forests **12**(4), 468.

Gemmeke, J., Ellis, D., Freedman, D., Jansen, A., Lawrence, W., Moore, R. C., Plakal, M., and Ritter, M. (**2017**). "AudioSet: An ontology and human-labeled dataset for audio events," in *Proceedings of ICASSP*, pp. 776–780.

Gong, Y., Chung, Y. A., and Glass, J. (**2021**). "AST: Audio Spectrogram Transformer," in *Proceedings of INTERSPEECH*, pp. 571–575.

Gong, Y., Cui, C., Cai, M., Dong, Z., Zhao, Z., and Wang, A. (**2022**). "Residents' preferences to multiple sound sources in urban park: Integrating soundscape measurements and semantic differences," Forests **13**(11), 1754.

Hanusz, Z., Tarasinska, J., and Zielinski, W. (**2016**). "Shapiro–Wilk test with known mean," REVSTAT-Stat. J. **14**(1), 89–100.

Hong, J. Y., and Jeon, J. Y. (**2015**). "Influence of urban contexts on soundscape perceptions: A structural equation modeling approach," Landscape Urban Plann. **141**, 78–87.

Hou, Y. (**2023**). "AI-Soundscape," https://github.com/Yuanbo2020/AI-Soundscape (Last viewed 9/11/2023).

Hou, Y., and Botteldooren, D. (**2022**). "Event-related data conditioning for acoustic event classification," in *Proceedings of INTERSPEECH*, pp. 1561–1565.

Hou, Y., Kang, B., Van Hauwermeiren, W., and Botteldooren, D. (**2022a**). "Relation-guided acoustic scene classification aided with event embeddings," in *Proceedings of International Joint Conference on Neural Networks*, pp. 1–8.

Hou, Y., Liu, Z., Kang, B., Wang, Y., and Botteldooren, D. (**2022b**). "CT-SAT: Contextual Transformer for Sequential Audio Tagging," in *Proceedings of INTERSPEECH*, pp. 4147–4151.

Iannace, G., Ciaburro, G., and Trematerra, A. (**2019**). "Wind turbine noise prediction using random forest regression," Machines **7**(4), 69.

Izonin, I., Tkachenko, R., Shakhovska, N., and Lotoshynska, N. (**2021**). "The additive input-doubling method based on the SVR with nonlinear kernels: Small data approach," Symmetry **13**(4), 612.

Kang, J., Aletta, F., Gjestland, T. T., Brown, L. A., Botteldooren, D., Schulte-Fortkamp, B., Lercher, P., van Kamp, I., Genuit, K., Fiebig, A., Bento Coelho, J. L., Maffei, L., and Lavia, L. (**2016**). "Ten questions on the soundscapes of the built environment," Build. Environ. **108**, 284–294.

Kingma, D. P., and Ba, J. (**2015**). "Adam: A method for stochastic optimization," in *Proceedings of International Conference on Learning Representations*.

Kliuchko, M., Heinonen-Guzejev, M., Vuust, P., Tervaniemi, M., and Brattico, E. (**2016**). "A window into the brain mechanisms associated with noise sensitivity," Sci. Rep. **6**(1), 39236.

Kong, Q., Cao, Y., Iqbal, T., Wang, Y., Wang, W., and Plumbley, M. D. (**2020**). "PANNs: Large-scale pretrained audio neural networks for audio pattern recognition," IEEE/ACM Trans. Audio. Speech. Lang. Process. **28**, 2880–2894.

Kruse, R., Mostaghim, S., Borgelt, C., Braune, C., and Steinbrecher, M. (**2022**). "Multi-layer perceptrons," in *Computational Intelligence: A Methodological Introduction* (Springer, Berlin), pp. 53–124.

Landström, U., Åkerlund, E., Kjellberg, A., and Tesarz, M. (**1995**). "Exposure levels, tonal components, and noise annoyance in working environments," Environ. Int. **21**(3), 265–275.

Lefèvre, M., Chaumond, A., Champelovier, P., Allemand, L. G., Lambert, J., Laumon, B., and Evrard, A. S. (**2020**). "Understanding the relationship between air traffic noise exposure and annoyance in populations living near airports in france," Environ. Int. **144**, 106058.

Lercher, P., and Schulte-Fortkamp, B. (**2003**). "The relevance of soundscape research to the assessment of noise annoyance at the community level," in *Proceedings of International Congress on Noise as a Public Health Problem*, pp. 225–231.
Li, Y., Liu, M., Drossos, K., and Virtanen, T. (**2020**). "Sound event detection via dilated convolutional recurrent neural networks," in *Proceedings of ICASSP*, pp. 286–290.
Li, Z., Hou, Y., Xie, X., Li, S., Zhang, L., Du, S., and Liu, W. (**2019**). "Multi-level attention model with deep scattering spectrum for acoustic scene classification," in *IEEE International Conference on Multimedia & Expo Workshops (ICMEW)*, pp. 396–401.
Ma, K. W., Mak, C. M., and Wong, H. M. (**2021**). "Effects of environmental sound quality on soundscape preference in a public urban space," Appl. Acoust. **171**, 107570.
Mackrill, J., Cain, R., and Jennings, P. (**2013**). "Experiencing the hospital ward soundscape: Towards a model," J. Environ. Psychol. **36**, 1–8.
Marchegiani, L., and Posner, I. (**2017**). "Leveraging the urban soundscape: Auditory perception for smart vehicles," in *IEEE International Conference on Robotics and Automation*, pp. 6547–6554.
Maristany, A., López, M. R., and Rivera, C. A. (**2016**). "Soundscape quality analysis by fuzzy logic: A field study in Cordoba, Argentina," Appl. Acoust. **111**, 106–115.
Mesaros, A., Heittola, T., Benetos, E., Foster, P., Lagrange, M., Virtanen, T., and Plumbley, M. D. (**2018a**). "Detection and classification of acoustic scenes and events: Outcome of the DCASE 2016 challenge," IEEE/ACM Trans. Audio. Speech. Lang. Process. **26**(2), 379–393.
Mesaros, A., Heittola, T., and Virtanen, T. (**2018b**). "Acoustic scene classification: An overview of DCASE 2017 challenge entries," in *International Workshop on Acoustic Signal Enhancement (IWAENC)*, pp. 411–415.
Mitchell, A., Brown, E., Deo, R., Hou, Y., Kirton-Wingate, J., Liang, J., Sheinkman, A., Soelistyo, C., Sood, H., and Wongprommoon, A. (**2023**). "Deep learning techniques for noise annoyance detection: Results from an intensive workshop at the Alan Turing Institute," J. Acoust. Soc. Am. **153**(3), A262.
Mitchell, A., Erfanian, M., Soelistyo, C., Oberman, T., Kang, J., Aldridge, R., Xue, J. H., and Aletta, F. (**2022**). "Effects of soundscape complexity on urban noise annoyance ratings: A large-scale online listening experiment," Int. J. Environ. Res. Public Health **19**(22), 14872.
Morrison, W. E., Haas, E. C., Shaffner, D. H., Garrett, E. S., and Fackler, J. C. (**2003**). "Noise, stress, and annoyance in a pediatric intensive care unit," Crit. Care Med. **31**(1), 113–119.
Mount, W. M., Tuček, D. C., and Abbass, H. A. (**2012**). "A psychophysiological analysis of weak annoyances in human computer interfaces," in *Proceedings of International Conference on Neural Information Processing*, pp. 202–209.
Mulimani, M., and Koolagudi, S. G. (**2018**). "Acoustic event classification using spectrogram features," in *TENCON 2018-2018 IEEE Region 10 Conference*, pp. 1460–1464.
Nering, K., Kowalska-Koczwara, A., and Stypuła, K. (**2020**). "Annoyance based vibro-acoustic comfort evaluation of as summation of stimuli annoyance in the context of human exposure to noise and vibration in buildings," Sustainability **12**(23), 9876.
Nilsson, M. E., and Berglund, B. (**2006**). "Soundscape quality in suburban green areas and city parks," Acta Acust. united Acust. **92**(6), 903–911.
Parascandolo, G., Huttunen, H., and Virtanen, T. (**2016**). "Recurrent neural networks for polyphonic sound event detection in real life recordings," in *2016 IEEE International Conference on Acoustics, Speech and Signal Processing (ICASSP)*, pp. 6440–6444.
Piczak, K. J. (**2015**). "ESC: Dataset for Environmental Sound Classification," in *Proceedings of Annual ACM Conference on Multimedia*, pp. 1015–1018.
Plakal, M., and Ellis, D. (**2023**). "YAMNet," https://github.com/tensorflow/models/tree/master/research/audioset/yamnet (Last viewed 9/11/2023).
Politis, A., Mesaros, A., Adavanne, S., Heittola, T., and Virtanen, T. (**2021**). "Overview and evaluation of sound event localization and detection in DCASE 2019," IEEE/ACM Trans. Audio. Speech. Lang. Process. **29**, 684–698.
Qu, F., Li, Z., Zhang, T., and Huang, W. (**2023**). "Soundscape and subjective factors affecting residents' evaluation of aircraft noise in the communities under flight routes," Front. Psychol. **14**, 1197820.
Raimbault, M., and Dubois, D. (**2005**). "Urban soundscapes: Experiences and knowledge," Cities **22**(5), 339–350.
Ren, J., Jiang, X., Yuan, J., and Magnenat-Thalmann, N. (**2017**). "Sound-event classification using robust texture features for robot hearing," IEEE Trans. Multimedia **19**(3), 447–458.
Salamon, J., Jacoby, C., and Bello, J. P. (**2014**). "A dataset and taxonomy for urban sound research," in *Proceedings of the ACM International Conference on Multimedia*, pp. 1041–1044.
Sun, K., De Coensel, B., Filipan, K., Aletta, F., Van Renterghem, T., De Pessemier, T., Joseph, W., and Botteldooren, D. (**2019**). "Classification of soundscapes of urban public open spaces," Landscape Urban Plann. **189**, 139–155.
Szychowska, M., Hafke-Dys, H., Preis, A., Kociński, J., and Kleka, P. (**2018**). "The influence of audio-visual interactions on the annoyance ratings for wind turbines," Appl. Acoust. **129**, 190–203.
Thorogood, M., Fan, J., and Pasquier, P. (**2016**). "Soundscape audio signal classification and segmentation using listeners perception of background and foreground sound," J. Audio Eng. Soc. **64**(7/8), 484–492.
Timmons, A. C., Han, S. C., Chaspari, T., Kim, Y., Narayanan, S., Duong, J. B., Simo Fiallo, N., and Margolin, G. (**2023**). "Relationship satisfaction, feelings of closeness and annoyance, and linkage in electrodermal activity," Emotion **23**, 1815.
Trotta, F. (**2020**). *Annoying Music in Everyday Life* (Bloomsbury Publishing, New York).
Tsaligopoulos, A., Kyvelou, S., Votsi, N., Karapostoli, A., Economou, C., and Matsinos, Y. G. (**2021**). "Revisiting the concept of quietness in the urban environment-towards ecosystems' health and human well-being," Int. J. Environ. Res. Public Health **18**(6), 3151.
Van Renterghem, T. (**2019**). "Towards explaining the positive effect of vegetation on the perception of environmental noise," Urban For. Urban Greening **40**, 133–144.
Vaswani, A., Shazeer, N., Parmar, N., Uszkoreit, J., Jones, L., Gomez, A. N., Kaiser, L., Polosukhin, I. (**2017**). "Attention is all you need," in *Proceedings of International Conference on Neural Information Processing Systems*, pp. 5998–6008.
Wallach, D., and Goffinet, B. (**1989**). "Mean squared error of prediction as a criterion for evaluating and comparing system models," Ecol. Modell. **44**(3-4), 299–306.
Xu, Y., Kong, Q., Huang, Q., Wang, W., and Plumbley, M. D. (**2017**). "Convolutional gated recurrent neural network incorporating spatial features for audio tagging," in *2017 International Joint Conference on Neural Networks (IJCNN)*, pp. 3461–3466.
Yilmazer, S., and Acun, V. (**2018**). "A grounded theory approach to assess indoor soundscape in historic religious spaces of Anatolian culture: A case study on Hacı Bayram mosque," Build. Acoust. **25**(2), 137–150.
Zhou, H., Shu, H., and Song, Y. (**2018**). "Using machine learning to predict noise-induced annoyance," in *IEEE Region 10 Conference*, pp. 0229–0234.